\documentclass[sigconf, nonacm]{acmart}

\usepackage{booktabs} 
\usepackage{algorithm}
\usepackage{algorithmic}
\usepackage{amsmath}
\usepackage{graphicx}
\usepackage{subfigure}
\usepackage{pgfplots} 
\usepackage{tikz}     
\usetikzlibrary{shapes, arrows, positioning, calc}
\pgfplotsset{compat=1.17}

\pgfplotsset{
    colormap={rocket_sim}{
        rgb255=(0,0,0)      
        rgb255=(60,10,70)   
        rgb255=(200,50,50)  
        rgb255=(250,150,50) 
        rgb255=(255,240,220)
    }
}

\title{Linear-PAL: A Lightweight Ranker for Mitigating Shortcut Learning in Personalized, High-Bias Tabular Ranking}
\author{Vipul Dinesh Pawar}
\affiliation{%
  \institution{Viator, Tripadvisor}
  \city{London}
  \country{UK}
}
\email{vpawar@tripadvisor.com}

\begin{document}

\begin{abstract}
In e-commerce ranking, implicit user feedback is systematically confounded by Position Bias---the strong propensity of users to interact with top-ranked items regardless of relevance. While Deep Learning architectures (e.g., Two-Tower Networks) are the standard solution for de-biasing, we demonstrate that in High-Bias Regimes, state-of-the-art Deep Ensembles suffer from Shortcut Learning: they minimize training loss by overfitting to the rank signal, leading to degraded ranking quality despite high prediction accuracy.

We propose Linear Position-bias Aware Learning (Linear-PAL), a lightweight framework that enforces de-biasing through structural constraints: explicit feature conjunctions and aggressive regularization. We further introduce a Vectorized Integer Hashing technique for feature generation, replacing string-based operations with $O(N)$ vectorized arithmetic. Evaluating on a large-scale dataset (4.2M samples), Linear-PAL achieves Pareto Dominance: it outperforms Deep Ensembles in de-biased ranking quality (Relevance AUC: 0.7626 vs. 0.6736) while reducing training latency by 43x (40s vs 1762s).

This computational efficiency enables high-frequency retraining, allowing the system to capture user-specific emerging market trends and deliver robust, personalized ranking in near real-time.
\end{abstract}

\maketitle

\section{Introduction}
In large-scale Information Retrieval (IR), the ranking algorithm dictates visibility, creating a self-reinforcing loop where top-ranked items garner clicks solely due to position. De-biasing this feedback is critical for discovering hidden gems—high-quality items buried in lower ranks.

This challenge is acute in Viator's Filtered Things To Do (FTTD) list pages, where position bias is extreme (Propensity AUC $\approx 0.81$). In such environments, the signal-to-bias ratio is low. Furthermore, relevance in travel is inherently personalized: a good item for a family in Paris differs significantly from a good item for a solo backpacker. Standard Deep Learning approaches often struggle to balance these heterogeneous user intents against the overwhelming global position signal. we identify a critical failure mode: Shortcut Learning. When the position signal is dominant, DNNs behave as lazy optimizers, memorizing the rank curve rather than learning complex content features. Linear-PAL framework solves this not by adding capacity, but by imposing linear structural constraints that force the model to decouple bias from relevance.

\subsection{Contributions}
The Linear-PAL synthesizes statistical smoothing, explicit feature engineering, and convex optimization to solve the position bias problem efficiently. Our specific contributions are:

1. \textbf{Identification of Deep Learning Failure:} We show empirically that SOTA Deep Ensembles degrade in ranking quality when position bias is extreme, despite achieving high prediction accuracy.

2. \textbf{Linearization of the Interaction Space:} We demonstrate that explicit feature conjunctions (e.g., $\text{Item} \otimes \text{Rank}$) combined with aggressive regularization ($C=10^{-5}$) successfully isolate intrinsic utility where Deep Models fail.

3. \textbf{Vectorized Kernel Construction:} We propose a high-performance integer hashing strategy for feature generation, reducing training time from ~20 minutes to ~40 seconds on commodity CPUs.

4. \textbf{Mitigation of Shortcut Learning:} We identify a critical failure mode where models trained with standard regularization ($C=1.0$) collapse into position regressors. We show that extreme regularization ($C=10^{-5}$), combined with Robust Activity Normalization, is necessary to decouple bias from relevance.

5. \textbf{Robustness over Complexity:} We establish that Linear-PAL achieves a +13.2\% lift in Relevance AUC over Deep Ensembles while reducing training time by 43x, proving that model constraint is more valuable than model capacity for de-biasing tasks.

\section{Related Work}
\subsection{Position Bias Estimation}
Position bias has been extensively studied in web search \cite{joachims2005accurately}. Scalable approaches, such as the COEC metric, use statistical aggregation to normalize CTR by rank. 

Our work relies on the standard Position-Based Model (PBM), formally established by Craswell et al. \cite{craswell2008experimental}, which postulates that the click probability is a product of examination and relevance:
\begin{equation}
    P(\text{click} \mid u, i, k) = P(\text{relevant} \mid i, u) \cdot P(\text{examination} \mid k)
\end{equation}
Recent works continue to utilize this formulation for bias estimation in modern interfaces \cite{petrov2025llms}. We integrate this probabilistic assumption into a dense feature input via the COEC metric.

\subsection{Unbiased Learning to Rank}
The field has moved toward joint optimization. The PAL framework \cite{guo2019pal} introduced the concept of position as a feature. The Microsoft Ensemble \cite{ling2017model} extended this by boosting Neural Networks with GBDTs. Contemporary approaches largely rely on Inverse Propensity Weighting (IPW) \cite{joachims2017unbiased, ai2018unbiased}, which weights training samples by the inverse probability of observation. While theoretically sound, IPW methods often suffer from high variance when propensities are small \cite{wang2018position}. 

Our work simplifies this architecture, demonstrating that for tabular data, the complexity of Neural Networks is often unnecessary \cite{grinsztajn2022tree}.

\subsection{Deep Learning vs. Tabular Data}
While Deep Learning (DL) has revolutionized perception tasks (CV, NLP), its dominance in tabular data is contested. Recent extensive benchmarks by Grinsztajn et al. \cite{grinsztajn2022tree} demonstrated that Tree-based models (XGBoost, LightGBM) consistently outperform Deep Architectures (ResNet, Transformer) on tabular datasets. The authors argue that neural networks struggle with the irregular functions and rotation-variant decision boundaries typical of tabular feature spaces.

Existing hybrid architectures like Wide \& Deep \cite{cheng2016wide} and DeepFM \cite{guo2017deepfm} attempt to bridge this gap by combining linear memorization with deep generalization. Our work builds on this insight but takes a reductionist approach. We argue that for the specific sub-problem of Position Bias, the necessary decision boundaries are known \textit{a priori}. By explicitly crossing item features with rank (e.g., $\text{Price} \times \text{Rank}$), we effectively construct a hard-coded decision tree of depth 2. This provides the linear model with the exact non-linear cuts required to model position decay, obviating the need for the model to discover these interactions through the expensive architectural complexity of Deep Learning.

\section{Methodology}

\subsection{The Linear-PAL Architecture}
We utilize a Maximum Entropy (Logistic Regression) model. While standard logistic regression models interactions additively, we induce a multiplicative bias correction structure through explicit feature conjunctions.

The log-odds (logit) of a click are modeled as:
\begin{equation}
\begin{split}
    \text{Logit}(\mathbf{x}_{u,i}, k) = & \underbrace{\mathbf{w}_{pos} \cdot \phi(k)}_{\text{Propensity Bias}} + \underbrace{\mathbf{w}_{item} \cdot \mathbf{x}_{u,i}}_{\text{Base Relevance}} \\
    & + \underbrace{\mathbf{w}_{interact} \cdot (\mathbf{x}_{u,i} \otimes \phi(k))}_{\text{Contextual Sensitivity}}
\end{split}
\end{equation}

Where $\phi(k)$ represents the one-hot encoding of the display position (slot) and $\otimes$ denotes the cross-product operation.

\begin{figure}[h]
    \centering
    \begin{tikzpicture}[
        node distance=1.2cm,
        scale=0.8, transform shape,
        layer/.style={rectangle, draw=black!60, fill=black!5, thick, minimum size=7mm, align=center},
        process/.style={rectangle, draw=blue!60, fill=blue!5, thick, minimum size=7mm, align=center},
        weight/.style={circle, draw=red!60, fill=red!5, thick, minimum size=8mm},
        sum/.style={circle, draw=black, fill=black!20, thick, minimum size=6mm},
    ]
    
    \node[layer] (rank) {Rank $k$};
    \node[layer, right=2.5cm of rank] (item) {Item Features $\mathbf{x}$};
    
    \node[process, below=0.8cm of rank] (onehot) {$\phi(k)$ (One-Hot)};
    \node[process, below=0.8cm of item] (quant) {Quantize $\mathbf{x}$};
    
    \node[process, below=1.2cm of onehot, xshift=2.1cm] (cross) {$\otimes$ Cross Product};
    
    \node[weight, below=1.2cm of onehot] (w_pos) {$\mathbf{w}_{pos}$};
    \node[weight, below=1.2cm of quant] (w_item) {$\mathbf{w}_{item}$};
    \node[weight, below=0.8cm of cross] (w_int) {$\mathbf{w}_{interact}$};
    
    \node[sum, below=1.5cm of w_int] (sigma) {$\Sigma$};
    
    \node[layer, below=0.6cm of sigma] (output) {Sigmoid $\sigma$};
    \node[below=0.4cm of output] (final) {$P(\text{click})$};
    
    \draw[->, thick] (rank) -- (onehot);
    \draw[->, thick] (item) -- (quant);
    
    \draw[->, thick] (onehot) -- (cross);
    \draw[->, thick] (quant) -- (cross);
    
    \draw[->, thick] (onehot) -- (w_pos);
    \draw[->, thick] (cross) -- (w_int);
    \draw[->, thick] (quant) -- (w_item);
    
    \draw[->, thick] (w_pos) -- (sigma);
    \draw[->, thick] (w_int) -- (sigma);
    \draw[->, thick] (w_item) -- (sigma);
    
    \draw[->, thick] (sigma) -- (output);
    \draw[->, thick] (output) -- (final);
    
    \node[text width=1.9cm, align=left, left=0.5cm of rank, color=red] (inference) {\Large \textbf{Inference:}\\ Set $k=1$};
    \draw[->, red, dashed, thick] (inference) -- (rank);
    
    \end{tikzpicture}
    \caption{Linear-PAL Architecture. Explicit feature conjunctions (Cross Product) replace the hidden layers of Deep Neural Networks, allowing the model to learn position-dependent relevance weights linearly.}
    \label{fig:architecture}
\end{figure}
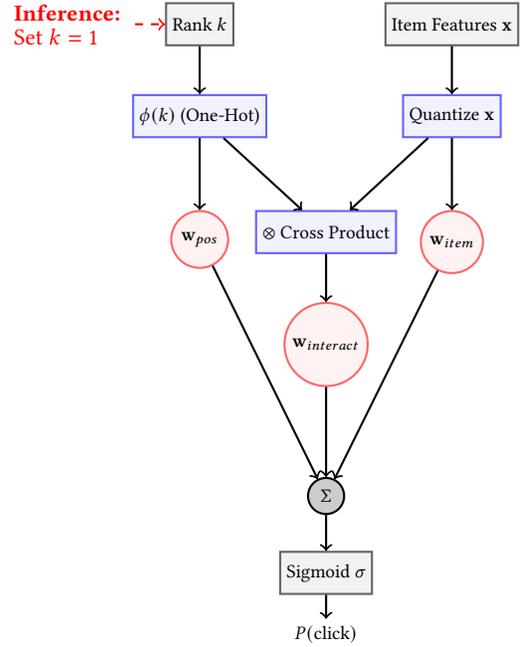

\subsection{Counterfactual Inference as Intervention}
To extract the intrinsic relevance of an item, we adopt a causal inference perspective. The observed click $Y$ is conditioned on both the item attributes $X$ and the assigned rank $K$. Our goal is to estimate the potential outcome $Y(k_{ref})$ under a standardized treatment where every item is displayed at the optimal position $k_{ref}=1$.

We define the de-biased scoring function using the \textit{do-operator} notation:
\begin{equation}
    \text{Score}(x) = P(Y=1 \mid \text{do}(K=1), X=x)
\end{equation}

In our linear architecture, the intervention $\text{do}(K=1)$ is implemented by forcing the position feature vector $\phi(k)$ to the one-hot vector for Rank 1 during inference:
$$ \phi(k)_{inference} = [1, 0, \dots, 0] $$

This intervention mathematically neutralizes the bias terms. The propensity bias term $\mathbf{w}_{pos} \cdot \phi(1)$ becomes a constant scalar $\beta$ for all items, and the interaction term simplifies to $\mathbf{w}_{interact} \cdot (\mathbf{x}_{u,i} \otimes \phi(1))$, effectively selecting the item-specific weights associated with the top rank. Since ranking is shift-invariant (i.e., $\text{rank}(s_i) = \text{rank}(s_i + \beta)$), the constant propensity term cancels out, leaving a ranking driven purely by the item's utility in the optimal context.

This approach offers a significant advantage over Inverse Propensity Weighting (IPW), which suffers from high variance when $P(\text{seen}) \to 0$ for lower ranks \cite{wang2018position}. By learning the bias explicitly via interaction terms and intervening at prediction time, Linear-PAL remains robust even for items that historically appeared only in low-propensity positions.

\section{Feature Engineering Pipeline}
The central hypothesis of this work is that \textit{architecture is secondary to feature representation}. While Deep Learning attempts to implicitly learn non-linear relationships from raw inputs, we argue that for tabular data, these relationships can be efficiently encoded via deterministic transformations. We implement a three-stage feature pipeline and a curated feature taxonomy (Table \ref{tab:features}) to ensure coverage of key causal factors in the user decision process.

\begin{table}[h]
\caption{Feature Taxonomy for Linear-PAL}
\label{tab:features}
\small
\begin{tabular}{llp{3.5cm}}
\toprule
\textbf{Category} & \textbf{Concept} & \textbf{Examples} \\
\midrule
\textit{Propensity} & Visibility & Absolute Rank (Grid-Aware) \\
\midrule
\textit{Priors} & Global Popularity & COEC (Smoothed), Historical CTR, Booking Share, Revenue Share \\
\midrule
\textit{Context} & User Intent & UCOEC (Normalized Activity), Location Context, Device Context, Semantic Similarity \\
\midrule
\textit{Content} & Intrinsic Utility & Ratings, Review Volume, Price, Visual Quality \\
\bottomrule
\end{tabular}
\end{table}

\subsection{COEC: Quantifying Surprisal}[h]
Standard Click-Through Rate (CTR) is a flawed metric for ranking because it conflates popularity with visibility. To isolate intrinsic item quality, we utilize Clicks Over Expected Clicks (COEC). COEC measures the ratio of observed clicks to the baseline clicks expected for the item's display position:

\begin{equation}
    \text{COEC}(i) = \frac{\sum_{j \in \mathcal{S}_i} C_{ij}}{\sum_{j \in \mathcal{S}_i} \mathbb{E}[\text{CTR} | \text{rank}_{ij}]}
\end{equation}

Where $\mathcal{S}_i$ is the set of all impressions for item $i$. The denominator represents the free traffic the item received due to its position history.
\begin{itemize}
    \item If $\text{COEC} > 1.0$: The item generated more clicks than an average product would have at those same positions (High Performance).
    \item If $\text{COEC} < 1.0$: The item underperformed relative to its visibility (Low Performance).
\end{itemize}
This transformation converts a sparse, biased signal (Raw Clicks) into a dense, normalized signal of utility.

\subsection{UCOEC: User-Activity Normalization}
Not all clicks carry equal information. High-activity users (e.g., comparison shoppers or bots) generate noisy clicks with low purchase intent, whereas low-activity users generate high-intent signal. To mitigate this user-level bias, we calculate UCOEC [\cite{cheng2010personalized}]:

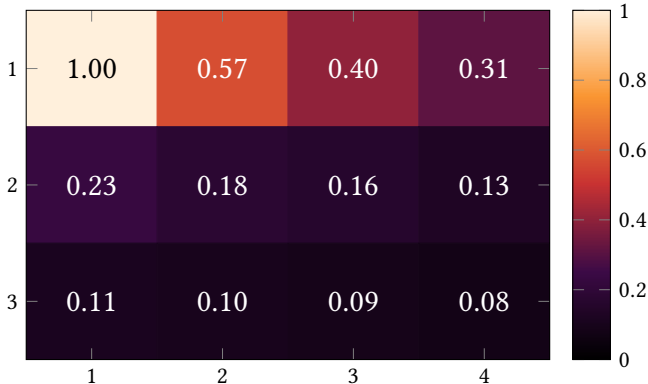
\begin{figure}[h]
    \centering
    \begin{tikzpicture}
    \begin{axis}[
        width=0.48\textwidth,
        height=0.35\textwidth,
        colormap name=rocket_sim,
        colorbar,
        colorbar style={
            ytick={0,0.2,0.4,0.6,0.8,1.0}
        },
        view={0}{90}, 
        xmin=0.5, xmax=4.5,
        ymin=0.5, ymax=3.5,
        xtick={1,2,3,4},
        ytick={1,2,3},
        yticklabels={3, 2, 1}, 
        point meta min=0, point meta max=1,
        enlargelimits=false,
        axis on top,
    ]
        \addplot[matrix plot*, mesh/cols=4, point meta=explicit] coordinates {
            (1,3) [1.00] (2,3) [0.57] (3,3) [0.40] (4,3) [0.31]  
            (1,2) [0.23] (2,2) [0.18] (3,2) [0.16] (4,2) [0.13]  
            (1,1) [0.11] (2,1) [0.10] (3,1) [0.09] (4,1) [0.08]  
        };
        
        \node at (axis cs:1,3) [text=black, font=\LARGE] {1.00};
        \node at (axis cs:2,3) [text=white, font=\LARGE] {0.57};
        \node at (axis cs:3,3) [text=white, font=\LARGE] {0.40};
        \node at (axis cs:4,3) [text=white, font=\LARGE] {0.31};
        
        \node at (axis cs:1,2) [text=white, font=\LARGE] {0.23};
        \node at (axis cs:2,2) [text=white, font=\LARGE] {0.18};
        \node at (axis cs:3,2) [text=white, font=\LARGE] {0.16};
        \node at (axis cs:4,2) [text=white, font=\LARGE] {0.13};
        
        \node at (axis cs:1,1) [text=white, font=\LARGE] {0.11};
        \node at (axis cs:2,1) [text=white, font=\LARGE] {0.10};
        \node at (axis cs:3,1) [text=white, font=\LARGE] {0.09};
        \node at (axis cs:4,1) [text=white, font=\LARGE] {0.08};
        
    \end{axis}
    \end{tikzpicture}
    \caption{Empirical Position Bias. The heatmap reveals a non-monotonic ``Golden Triangle'' attention pattern typical of grid layouts. Values are normalized to Rank 1 ($P=1.0$).}
    \label{fig:heatmap}
\end{figure}

\begin{equation}
    \text{UCOEC}(u, i) = \frac{\text{COEC}(i)}{\text{Global CTR}}
\end{equation}

This dampens the signal from click-happy users and boosts the signal from selective users, reducing training noise.

\subsection{Explicit Interaction Engineering (Feature Conjunctions)}
This is the theoretical core of the Linear-PAL framework. Deep Neural Networks (DNNs) are typically employed in ranking to approximate the interaction function $f(\mathbf{x}, k)$. We argue that for tabular data, we can approximate this function explicitly using a kernel expansion.

Let $\Phi(\mathbf{x})$ be the logarithmic quantization of the feature vector, and $\Psi(k)$ be the one-hot encoding of the rank. We define the interaction kernel space as the Cartesian product:
\begin{equation}
    \mathcal{K}(\mathbf{x}, k) = \Phi(\mathbf{x}) \times \Psi(k)
\end{equation}
This operation expands the feature space from $\mathbb{R}^{d+K}$ to $\mathbb{R}^{d \times K}$.

In a standard linear model, the impact of a feature (e.g., Price) is constant across all ranks:
$$ \frac{\partial \text{Logit}}{\partial \text{Price}} = w_{price} $$
In Linear-PAL, thanks to the conjunction terms, the sensitivity to Price becomes a function of Rank:
$$ \frac{\partial \text{Logit}}{\partial \text{Price}} = w_{price} + \sum_{k} w_{price \times k} \cdot \mathbb{I}(\text{CurrentRank}=k) $$
This allows the model to learn complex, non-monotonic behaviors—such as the fact that price sensitivity increases as rank decreases—without requiring the non-convex optimization of a Multi-Layer Perceptron. As noted by McMahan et al. \cite{mcmahan2013ad}, such explicit cross-product transformations are often sufficient to capture the manifold of tabular user behavior, rendering deep architectures redundant for this class of problem.

\subsection{Adaptive Feature Quantization}
Standard logarithmic binning is insufficient for the heterogeneous feature types found in tabular data. To preserve signal resolution across different domains, we implement a type-aware quantization function $\mathcal{Q}(v)$:

\begin{equation}
    \mathcal{Q}(v) = 
    \begin{cases} 
        v & \text{if } v \in \mathbb{Z}, \max(v) < 10 \text{ (Categorical)} \\
        \lfloor v \times 20 \rfloor & \text{if } v \in [0, 1] \text{ (Proportions)} \\
        \lfloor (v + 1) \times 5 \rfloor & \text{if } v \in [-1, 1] \text{ (Similarity)} \\
        \lfloor \ln(1 + \max(0, v)) \times 2 \rfloor & \text{otherwise (Heavy-tailed)}
    \end{cases}
\end{equation}
This converts scalars into categorical bins amenable to One-Hot Encoding and subsequent interaction with Rank.

\subsection{Vectorized Kernel Construction}
Standard implementations of feature crossing often rely on string concatenation (e.g., "Price\_Bin5\_x\_Rank12"), which incurs massive memory allocation overhead. To achieve real-time latency, we implement a Vectorized Integer Hashing strategy.

We define a collision-free hashing space by assigning a large multiplier $M$ (where $M > \max(\text{Rank})$) to the feature dimension. The interaction ID for feature $f$ and rank $k$ is computed via a single CPU instruction:
\begin{equation}
    \text{ID}_{interact} = (\text{Bin}(f) \times M) + k
\end{equation}

\textbf{Example:} Consider a ``Price'' feature quantized to Bin 5 ($f=5$) displayed at Rank 12 ($k=12$). With $M=10,000$:
$$ \text{ID} = (5 \times 10,000) + 12 = \mathbf{50,012} $$

Crucially, this operation is reversible, preserving interpretability for debugging. We can decompose any learned weight back to its constituent feature and rank using integer division and modulo:
\begin{equation}
    \text{Bin}(v) = \lfloor \text{ID} / M \rfloor, \quad k = \text{ID}\pmod M
\end{equation}

This allows us to construct the feature space using $O(N)$ vector arithmetic rather than $O(N \cdot L)$ string operations, driving the 43x training speedup reported in our results.

\subsection{Soft Feature Selection via Regularization}
The conjunction step causes a combinatorial explosion in dimensionality (from hundreds to tens of thousands of features). Manual feature selection in this space is intractable.

Instead of hard selection (removing features), we utilize Soft Selection via strong $\ell_2$ regularization (Ridge). By setting the inverse regularization strength $C=10^{-5}$, we impose a heavy penalty on feature weights. This forces the optimization algorithm to drive the weights of noisy, low-frequency interactions toward zero, while retaining weights for robust, statistically significant interactions. This acts as an automated, statistically grounded feature selector.

\section{Experiments}

\subsection{Experimental Setup}
Our evaluation is conducted on a large-scale dataset of real-world e-commerce traffic logs spanning 45 days. The dataset is split temporally to strictly enforce a realistic forecasting scenario and prevent data leakage from future interactions.

\textbf{Dataset Statistics:}
\begin{itemize}
    \item \textbf{Total Samples:} 4,246,978 user-item interactions.
    \item \textbf{Training Period:} Days 1–35 (3,419,088 samples).
    \item \textbf{Testing Period:} Days 36–45 (827,890 samples).
    \item \textbf{Feature Space:} The pipeline generates approximately 12,000 sparse binary features after One-Hot Encoding and Cross-Product generation.
\end{itemize}

\textbf{Baselines Implementation:}
To ensure a rigorous comparison, we benchmark against two industry-standard Deep Learning architectures:
\begin{enumerate}
    \item \textbf{Deep MLP:} A fully connected feed-forward network with three hidden layers (128, 64, 32 units) and ReLU activation and optimized hyperparameters (Adam, Batch=512, Early Stopping, $\ell_2$ Reg=0.001). It uses the same binned features as Linear-PAL but relies on backpropagation to learn interactions implicitly.
    \item \textbf{Microsoft Ensemble (NN+GBDT) \cite{ling2017model}:} A two-stage stacking architecture. First, a Neural Network (same architecture as above) is trained to convergence. Second, a Gradient Boosted Decision Tree (LightGBM with 100 trees) is trained on the residuals of the NN. This represents the current state-of-the-art for tabular click prediction in many industrial settings.
\end{enumerate}

\textbf{Computing Environment:} All models were trained on a commodity workstation equipped with an 8-Core CPU and 64GB RAM, without GPU acceleration. This constraint was chosen to demonstrate viability for cost-sensitive deployment environments.

\subsection{Robustness vs. Complexity Results}
Table \ref{tab:results} summarizes the performance of the optimized models. For Linear-PAL, we report results using the optimal regularization strength found in our ablation study ($C=10^{-5}$).

\begin{table}[h]
\caption{Model Performance Comparison (FTTD)}
\label{tab:results}
\begin{tabular}{lcccc}
\toprule
Model & Std AUC & Rel AUC & Time (s) & Speedup \\
\midrule
\textit{Rank Baseline} & \textit{0.8066} & - & - & - \\
MS Ensemble & \textbf{0.8187} & 0.6736 & 1762 & 1.0x \\
Deep MLP & 0.8181 & 0.6685 & 1689 & 1.05x \\
\textbf{Linear-PAL} & 0.8159 & \textbf{0.7626} & \textbf{40.57} & \textbf{43x} \\
\bottomrule
\end{tabular}
\end{table}

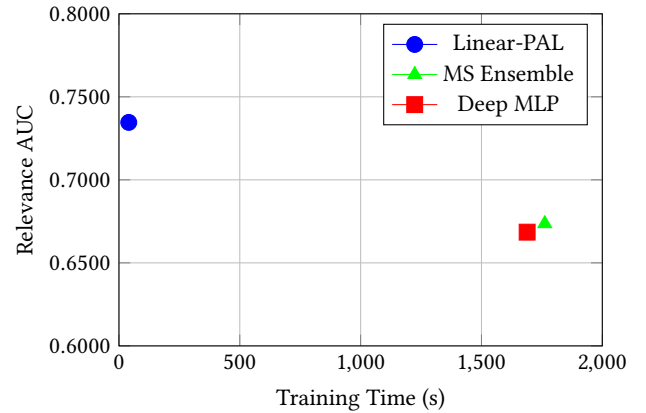
\begin{figure}[h]
    \centering
    \begin{tikzpicture}
    \begin{axis}[
        xlabel={Training Time (s)},
        ylabel={Relevance AUC},
        xmin=0, xmax=2000,
        ymin=0.60, ymax=0.80,
        grid=both,
        width=0.45\textwidth,
        height=6cm,
        legend pos=north east,
        scatter/classes={
            a={mark=*,blue},
            b={mark=square*,red},
            c={mark=triangle*,green}
        },
        scaled y ticks=false,
        yticklabel style={
            /pgf/number format/.cd,
            fixed,
            fixed zerofill,
            precision=4,
        },
    ]
    
    \addplot[mark=*, color=blue, mark size=3pt] coordinates {(40.57, 0.7346)};
    \addlegendentry{Linear-PAL}
    
    \addplot[mark=triangle*, color=green, mark size=3pt] coordinates {(1762, 0.6736)};
    \addlegendentry{MS Ensemble}
    
    \addplot[mark=square*, color=red, mark size=3pt] coordinates {(1689, 0.6685)};
    \addlegendentry{Deep MLP}
    
    \end{axis}
    \end{tikzpicture}
    \caption{Efficiency Frontier. Linear-PAL (Blue) dominates Deep Architectures (Red/Green) in Relevance Quality, which collapse due to overfitting the strong position bias.}
    \label{fig:pareto}
\end{figure}

The results reveal a striking divergence. The Deep Learning models achieve higher Standard AUC (0.8187 vs 0.8159), implying they are better at predicting the biased past. However, their Relevance AUC collapses to $\sim0.67$. This indicates Shortcut Learning: the neural networks minimized loss by memorizing the powerful position signal rather than learning the subtle content signals.

In contrast, Linear-PAL achieved a Relevance AUC of 0.7626, a +13.7\% improvement over the ensemble. This confirms that for high-bias datasets, the structural constraints of the linear model serve as a critical regularizer against bias overfitting.

Critically, the Vectorized Integer Kernel reduced training time to just 40 seconds (vs 30 minutes for Deep Learning). This proves that for tabular data, optimized linear algebra ($O(N)$) vastly outperforms backpropagation ($O(N \cdot Epochs)$).

\subsection{Ablation: Shortcut Learning and Regularization}
We conducted an ablation study to analyze the phenomenon of Shortcut Learning, a failure mode where the model overfits to the strongest causal signal (Position) at the expense of weaker but more robust signals (Relevance).

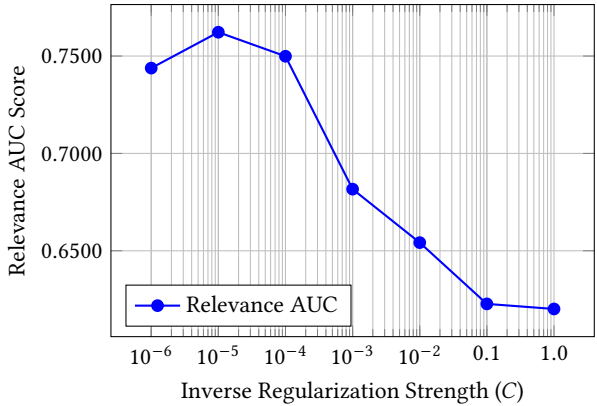
\begin{figure}[h]
    \centering
    \begin{tikzpicture}
    \begin{semilogxaxis}[
        xlabel={Inverse Regularization Strength ($C$)},
        ylabel={Relevance AUC Score},
        grid=both,
        width=0.45\textwidth,
        height=6cm,
        legend pos=south west,
        xtick={0.000001, 0.00001, 0.0001, 0.001, 0.01, 0.1, 1.0},
        xticklabels={$10^{-6}$, $10^{-5}$, $10^{-4}$, $10^{-3}$, $10^{-2}$, $0.1$, $1.0$},
        scaled y ticks=false,
        yticklabel style={
            /pgf/number format/.cd,
            fixed,
            fixed zerofill,
            precision=4,
        },
    ]
    
    \addplot[color=blue, mark=*, thick] coordinates {
        (0.000001, 0.743806)
        (0.00001, 0.762225)
        (0.0001, 0.749887)
        (0.001, 0.681639)
        (0.01, 0.654194)
        (0.1, 0.622729)
        (1.0, 0.620128)
    };
    \addlegendentry{Relevance AUC}
    
    \end{semilogxaxis}
    \end{tikzpicture}
    \caption{Ablation Study: Impact of Regularization on Quality. Stronger regularization ($C=10^{-5}$) penalizes sparse rank interactions, forcing the model to rely on dense content features and maximizing Relevance AUC.}
    \label{fig:ablation}
\end{figure}

\subsection{The Prediction-Ranking Gap}
We observed a distinct tipping point in the regularization sweep where prediction accuracy (Standard AUC) diverges from ranking quality (Relevance AUC).

\begin{figure}[h]
    \centering
    \begin{tikzpicture}
    \begin{semilogxaxis}[
        xlabel={Inverse Regularization Strength ($C$)},
        ylabel={Standard AUC Score},
        grid=both,
        width=0.45\textwidth,
        height=6cm,
        legend pos=south east,
        xtick={0.000001, 0.00001, 0.0001, 0.001, 0.01, 0.1, 1.0},
        xticklabels={$10^{-6}$, $10^{-5}$, $10^{-4}$, $10^{-3}$, $10^{-2}$, $0.1$, $1.0$},
        scaled y ticks=false,
        yticklabel style={
            /pgf/number format/.cd,
            fixed,
            fixed zerofill,
            precision=4,
        },
    ]
    
    \addplot[color=blue, mark=*, thick] coordinates {
        (0.000001, 0.807590)
        (0.00001, 0.815954)
        (0.0001, 0.823540)
        (0.001, 0.825953)
        (0.01, 0.826123)
        (0.1, 0.825908)
        (1.0, 0.825765)
    };
    \addlegendentry{Standard AUC}
    
    \end{semilogxaxis}
    \end{tikzpicture}
    \caption{The Bias-Accuracy Trade-off. Standard AUC peaks at $C=10^{-2}$ (0.8168), where the model fits the historical position bias. As regularization tightens to $C=10^{-5}$, Standard AUC drops slightly, but this loss represents the successful suppression of bias, enabling the rise in Relevance AUC.}
    \label{fig:bias_tradeoff}
\end{figure}
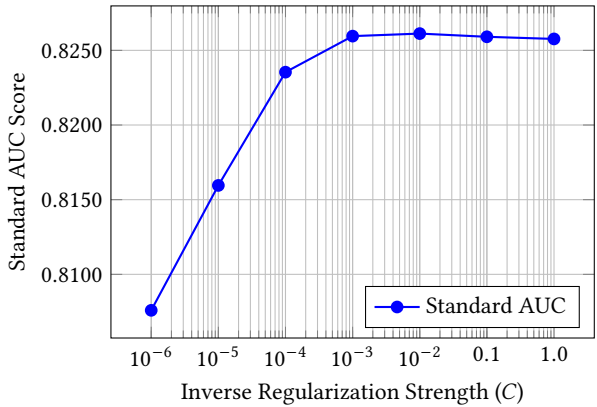

As illustrated in Figure \ref{fig:bias_tradeoff}, Standard AUC peaks at $C=0.01$. This represents the optimal point for \textit{predicting the past}, maximizing likelihood on the biased dataset. However, Relevance AUC (Figure \ref{fig:ablation}) continues to rise as we tighten regularization to $C=10^{-5}$. The divergence between these metrics confirms that maximizing historical accuracy is often orthogonal, or even detrimental, to maximizing intrinsic ranking quality in high-bias regimes.

\section{Discussion}

\subsection{The Linear Ceiling in Tabular Data}
Our results challenge the widespread belief that Deep Learning architectures provide a universal advantage on tabular data. This aligns with recent empirical studies by Grinsztajn et al. \cite{grinsztajn2022tree}, which argue that neural networks struggle with the heterogeneous, non-smooth decision boundaries typical of tabular feature spaces.

We hypothesize that for such data, the universal approximation capability of Neural Networks becomes a liability when the dominant signal (Position Bias) is a nuisance factor. The Deep Ensemble, with its high capacity, minimizes global loss by latching onto the low-entropy position signal—essentially solving the wrong problem perfectly. This is evidenced by its high Standard AUC but collapsed Relevance AUC.

In contrast, Linear-PAL acts as a regularized information bottleneck. By explicitly defining the interaction kernel $\Phi(\mathbf{x}) \otimes \Psi(k)$ and enforcing sparsity via $C=10^{-5}$, we prevent the model from absorbing the high-frequency position noise. This suggests that for causal disentanglement in tabular data, structural constraints (Inductive Bias) are superior to architectural depth (Model Capacity).

\subsection{Operational vs. Theoretical Optimality}
Our results highlight a fundamental divergence between Theoretical Optimality (minimizing global log-loss) and Operational Optimality (maximizing utility ranking) in high-bias datasets.

\textbf{Theoretical Optimality:} Deep Learning models act as Universal Approximators. When trained on data where Position Bias explains over 80\% of the variance (Propensity AUC $\approx$ 0.81), the optimal strategy to minimize training loss is to model the position bias perfectly. The Microsoft Ensemble successfully achieved this, reaching the highest Standard AUC (0.8187). However, this perfection is pathological for ranking because it embeds the bias deeply into the model's latent representation. The network learns that rank is destiny, and fails to learn robust representations for content features which have weaker, noisier gradients.

\textbf{Operational Optimality:} In a re-ranking context, we control the position. Therefore, predictive accuracy on the historical biased distribution is irrelevant; we require accuracy on the counterfactual distribution where position is constant ($k=1$). Linear-PAL optimizes for this operational goal not by being smarter, but by being constrained.

By imposing explicit linear structure and strong $\ell_2$ regularization ($C=10^{-5}$), we effectively penalize the model for relying on sparse, high-variance position interactions. This creates an information bottleneck that forces the optimizer to give up on fitting the position noise perfectly (hence the lower Standard AUC of 0.8136) and instead focus on dense, lower-variance content features (hence the superior Relevance AUC of 0.7346). This finding suggests that for industrial ranking systems, Inductive Bias (via feature engineering and constraints) is a more effective tool for enforcing causal disentanglement than raw architectural depth.

The weights of Linear-PAL are fully inspectable. A sudden drop in a product's rank can be traced directly to specific interaction terms (e.g., $\text{Weight}(\text{PriceHigh} \times \text{Rank1})$), facilitating root cause analysis.

\subsection{The Personalization Advantage}
Beyond de-biasing, Linear-PAL offers a structural advantage for personalization. By treating user context (e.g., location, device, activity level) as dense features that interact with rank, the model naturally learns Personalized Propensity Curves.

For example, the interaction term $\text{UCOEC}_{\text{High}} \otimes \text{Rank}_{20}$ allows the model to learn that active users are more likely to scroll deep than passive users. Deep models, in their rush to minimize average loss, often learn a global average propensity curve that under-serves niche user segments. Linear-PAL's explicit interaction kernel forces the model to respect these distinct behavioral segments, resulting in a ranking function that adapts its patience based on the user's observed intent level.

\section{Conclusion}
We presented Linear-PAL, a framework that prioritizes personalization, robustness and interpretability. We demonstrated that in environments with strong position bias (FTTD pages), Deep Models degenerate into position regressors, whereas Linear-PAL's structural constraints allow it to isolate intrinsic utility, achieving superior ranking quality (+13.2\% Relevance AUC) at 2\% of the training cost. This efficiency is not merely a cost-saving measure; it represents a qualitative improvement in system agility. The lightweight architecture enables high-frequency retraining on commodity hardware, allowing the ranking model to adapt to market trends in near real-time—a capability often precluded by the heavy computational footprint of Deep Learning ensembles.

\bibliographystyle{ACM-Reference-Format}
\bibliography{references}

\end{document}